\documentclass[preprints,article,accept,pdftex]{Definitions/mdpi} 
\firstpage{1} 
\pubvolume{1}
\issuenum{1}
\articlenumber{0}
\pubyear{2022}
\copyrightyear{2022}
\datereceived{} 
\dateaccepted{} 
\datepublished{} 
\hreflink{https://doi.org/} % If needed use \linebreak
\usepackage{subfig}
\usepackage{stmaryrd}
\usepackage[ruled]{algorithm2e} 
\usepackage{tensor}
\usepackage{float}

\newcommand{\R}{\mathbb{R}}

\newcommand{\E}{\mathbb{E}}
\newcommand{\Pb}{\mathbb{P}}

\newcommand{\cF}{\mathcal{F}}

\DeclareMathOperator{\Vol}{Vol}
\DeclareMathOperator{\Cut}{Cut}

\usepackage{mathtools}
\DeclarePairedDelimiterX{\norm}[1]{\lVert}{\rVert}{#1}

\DeclareMathOperator*{\argmin}{arg\,min}

\newcommand{\reflemma}[1]{Lemma~\ref{#1}}
\newcommand{\refcor}[1]{Corollary~\ref{#1}}

\DeclareMathOperator{\diag}{diag}

\Title{Mean estimation on the diagonal of product manifolds}

\TitleCitation{Mean estimation on the diagonal of product manifolds}

\newcommand{\orcidauthorB}{0000-0001-6784-0328} % Add \orcidB{} behind the author's name

\Author{Mathias H\o jgaard Jensen $^{1,\dagger,\ddagger}$ and Stefan Sommer $^{1,\ddagger}$\orcidauthorB}

\AuthorNames{Mathias H\o jgaard Jensen and Stefan Sommer}

\AuthorCitation{Jensen, M.; Sommer, S.}
\address{%
$^{1}$ \quad Department of Computer Science, University of Copenhagen; \{matje,sommer\}@di.ku.dk}

\corres{Correspondence: sommer@di.ku.dk}

\firstnote{Current address: Universitetsparken 5, DK-2100, Copenhagen E, Denmark} 
\secondnote{These authors contributed equally to this work.}
\abstract{Computing sample means on Riemannian manifolds is typically computationally costly as exemplified by computation of the Fréchet mean which often requires finding minimizing geodesics to each data point for each step of an iterative optimization scheme. When closed-form expressions for geodesics are not available, this leads to a nested optimization problem that is costly to solve. The implied computational cost impacts applications in both geometric statistics and in geometric deep learning. The weighted diffusion mean offers an alternative to the weighted Fr\'echet mean. We show how the diffusion mean and the weighted diffusion mean can be estimated with a stochastic simulation scheme that does not require nested optimization. We achieve this by conditioning a Brownian motion in a product manifold to hit the diagonal at a predetermined time. We develop the theoretical foundation for the sampling-based mean estimation, we develop two simulation schemes, and we demonstrate the applicability of the method with examples of sampled means on two manifolds.}


\begin{document}

\section{Introduction}
The Euclidean expected value can be generalized to geometric spaces in several ways. Fréchet \cite{frechet1948elements} generalized the notion of mean values to arbitrary metric spaces as minimizers of the sum of squared distances. Fréchet's notion of mean values thereby naturally includes means on Riemannian manifolds. On manifolds without metric, for example, affine connection spaces, a notion of the mean can be defined by exponential barycenters, see e.g. \cite{arnaudon2005barycenters,pennec_barycentric_2018}. Recently, Hansen et al. \cite{hansen2021diffusion,hansen2021diffusiongeometric} introduced a probabilistic notion of a mean, the diffusion mean. The diffusion mean is defined as the most likely starting point of a Brownian motion given the observed data. The variance of the data is here modelled in the evaluation time $T>0$ of the Brownian motion, and Varadhan's asymptotic formula relating the heat kernel with the Riemannian distance relates the diffusion mean and the Fréchet mean in the $T\to 0$ limit.
 
Computing sample estimators of geometric means is often difficult in practice. For example, estimating the Fréchet mean often requires evaluating the distance to each sample point at each step of an iterative optimization to find the optimal value. When closed-form solutions of geodesics are not available, the distances are themselves evaluated by minimizing over curves ending at the data points, thus leading to a nested optimization problem. This is generally a challenge in geometric statistics, the statistical analysis of geometric data. However, it can pose an even greater challenge in geometric deep learning, where a weighted version of the Fréchet mean is used to define a generalization of the Euclidean convolution taking values in a manifold \cite{chakraborty2020manifoldnet}. As the mean appears in each layer of the network, closed-form geodesics is in practice required for its evaluation to be sufficiently efficient. 
\begin{figure}[t]
    \centering
    \subfloat{
        \def\svgwidth{0.45\textwidth}
        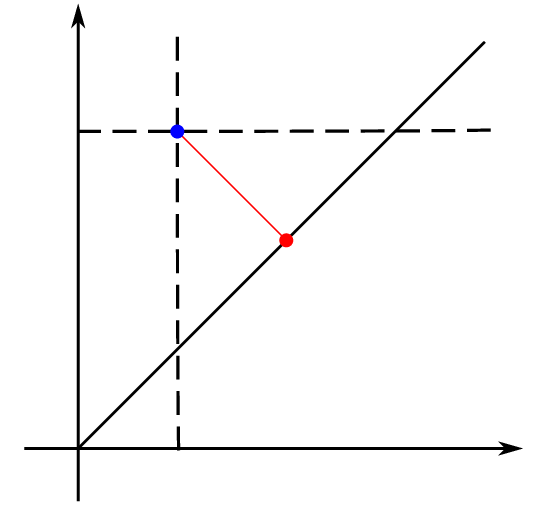
      }
    \subfloat{
        \includegraphics[width=0.45\linewidth]{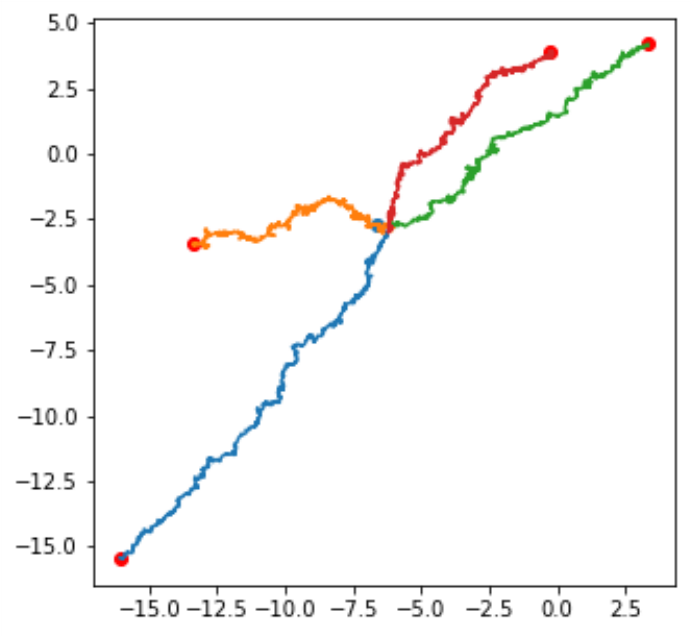}
      }
      \caption{(left) The mean estimator viewed as a projection onto the diagonal of a product manifold. Given a set $x_1,\dots, x_n \in M$, the tuple $(x_1,\dots, x_n)$ (blue dot) belongs to the product manifold $M\times \cdots \times M$. The mean estimator $\hat \mu$ can be identified with the projection of $(x_1,\dots,x_n)$ onto the diagonal $N$ (red dot). (right) Diffusion mean estimator in $\R^2$ using Brownian bridges conditioned on the diagonal. Here a Brownian bridge $X_t = (X_{1,t},\dots, X_{4,t})$ in $\R^8$ is conditioned on hitting the diagonal $N \subseteq \R^8$ at time $T>0$. The components $X_j$ each being two-dimensional processes are shown in the plot.}
\label{fig:intro}
\end{figure}

As an alternative to the weighted Fr\'echet mean, \cite{sommer2020horizontal} introduced a corresponding weighted version of the diffusion mean. 
Estimating the diffusion mean usually requires ability to evaluate the heat kernel making it often similarly computational difficult to estimate. However, \cite{sommer2020horizontal} also sketched a simulation based approach for estimating the (weighted) diffusion mean that avoids numerical optimization and estimation of the heat kernel. Here, a mean candidate is generated by simulating a single forward pass of a Brownian motion on a product manifold conditioned to hit the diagonal of the product space. The idea is sketched for samples in $\mathbb R^2$ in Figure~\ref{fig:intro}.

\subsection{Contribution}
In this paper, we present a comprehensive investigation of the simulation based mean sampling approach. We provide the necessary theoretical background and results for the construction, we present two separate simulation schemes, and we demonstrate how the schemes can be used to compute means on high-dimensional manifolds.

\section{Background}
We here outline the necessary concepts from Riemannian geometry, geometric statistics, stochastic analysis, and bridge sampling necessary for the sampling schemes presented later in the paper.

\subsection{Riemannian geometry}
A Riemannian metric $g$ on a $d$-dimensional differentiable manifold $M$ is a family of innner products $(g_p)_{p \in M}$ on each tangent space $T_p M$ varying smoothly in $p$. The Riemannian metric allows for geometric definitions of, e.g., length of curves, angles of intersections, and volumes on manifolds. A differentiable curve on $M$ is a map $\gamma \colon [0,1] \rightarrow M$ for which the time derivative $\gamma'(t)$ belongs to $T_{\gamma_t}M$, for each $t \in (0,1)$. The length of the differentiable curve can then be determined from the Riemannian metric by $L(\gamma) := \int_0^1 \sqrt{g_{\gamma_t}\left(\gamma'(t), \gamma'(t)\right)} dt = \int_0^1 \lVert \gamma'(t) \rVert_{\gamma_t} dt$. Let $p,q \in M$ and let $\Gamma$ be the set of differentiable curves joining $p$ and $q$, i.e., $\Gamma = \{\gamma \colon [0,1] \rightarrow M | \gamma(0) = p$ and $\gamma(1) =q\}$. The (Riemannian) distance between $p$ and $q$ is defined as $d(p,q) = \min_{\gamma \in \Gamma} L(\gamma)$. Minimizing curves are called geodesics.

A manifold can be parameterized using coordinate charts. The charts consist of open subsets of $M$ providing a global cover of $M$ such that each subset is diffeomorphic to an open subset of $\R^d$, or, equivalently, $\R^d$ itself. The exponential normal chart is often a convenient choice to parameterize a manifold for computational purposes. The exponential chart is related to the exponential map $\exp_p \colon T_pM \rightarrow M$ that for each $p\in M$ is given by $\exp_p(v) = \gamma_v(1)$, where $\gamma_v$ is the unique geodesic satisfying $\gamma_v(0)=p$ and $\gamma_v'(0) = v$. For each $p \in M$, the exponential map is a diffeomorphism from a star-shaped subset $V$ centered at the origin of $T_pM$ to its image $\exp_p(V) \subseteq M$, covering all of $M$ except for a subset of (Riemannian) volume measure zero, $\Cut(p)$, the cut-locus of $p$. The inverse map $\log_p \colon M\backslash \Cut(p) \rightarrow T_pM$ provides a local parameterization of $M$ due to the isomorphism between $T_pM$ and $\R^d$. For submanifolds $N\subseteq M$, the cut-locus $\Cut(N)$ is defined in a fashion similar to $\Cut(p)$, see e.g. \cite{thompson2015submanifold}.

Stochastic differential equations on manifolds are often conveniently expressed using the frame bundle $FM$, the fiber bundle which for each point $p \in M$ assigns a frame or basis for the tangent space $T_pM$, i.e., $FM$ consists of a collection of pairs $(p,u)$, where $u \colon \R^d \rightarrow T_pM$ is a linear isomorphism. We let $\pi$ denote the projection $\pi:FM\to M$. There exist a subbundle of $FM$ consisting of orthonormal frames called the orthonormal frame bundle $OM$. In this case, the map $u \colon \R^d \rightarrow T_pM$ is a linear isometry.

\subsection{Weighted Fréchet mean}
The Euclidean mean has three defining properties: The \textit{algebraic property} states the uniqueness of the arithmetic mean as the mean with residuals summing to zero, the \textit{geometric property} defines the mean as the point that minimizes the variance, and the \textit{probabilistic property} adheres to a maximum likelihood principle given an i.i.d. assumption on the observations (see also \cite[Chapter 2]{pennec_riemannian_2020}).
Direct generalization of the arithmetic mean to non-linear spaces is not possible due to the lack of vector space structure. However, the properties above allow giving candidate definitions of mean values in non-linear spaces.

The Fréchet mean \cite{frechet1948elements} uses the geometric property by generalizing the mean-squared distance minimization property to general metric spaces. Given a random variable $X$ on a metric space $(E,d)$, the Fréchet mean is defined by
    \begin{equation}\label{eq: frechet mean}
        \mu = \argmin_{p \in E} \E\left[d(p,X)^2 \right].
    \end{equation}
In the present context, the metric space is a Riemannian manifold $M$ with Riemannian distance function $d$. Given realizations $x_1,\dots, x_n \in M$ from a distribution on $M$, the estimator of the weighted Fréchet mean is defined as
    \begin{equation}\label{eq: weighted frechet mean}
        \hat \mu = \argmin_{p \in M} \sum_{i = 1}^n w_i d(p, x_i)^2,
    \end{equation}
where $w_1, \dots, w_n$ are the corresponding weights satisfying $w_i > 0$ and $\sum_i w_i = 1$. When the weights are identical, \eqref{eq: weighted frechet mean} is an estimator of the Fréchet mean. Throughout, we shall make no distinction between the estimator and the Fréchet mean and will refer to both as the Fréchet mean.

In \cite{pennec_riemannian_2006,chakraborty2020manifoldnet}, the weighted Fréchet mean was used to define a generalization of the Euclidean convolution to manifold-valued inputs. When closed-form solutions of geodesics are available, the weighted Fr\'echet mean can be estimated efficiently with a recursive algorithm, also denoted an inductive estimator \cite{chakraborty2020manifoldnet}.

\subsection{Weighted diffusion mean}
The diffusion mean \cite{hansen2021diffusion,hansen2021diffusiongeometric} was introduced as a geometric mean satisfying the probabilistic property of the Euclidean expected value, specifically as the starting point of a Brownian motion that is most likely given observed data. This results in the diffusion $t$-mean definition
    \begin{equation}
        \mu_t = \argmin_{p\in M} \E\left[ - \log p_{t}(p,X) \right],
    \end{equation}
where $p_t(\cdot, \cdot)$ denotes the transition density of a Brownian motion on $M$. Equivalently, $p_t$ denotes the solution to the heat equation $\partial u/ \partial t = \frac12 \Delta u$, where $\Delta$ denotes the Laplace-Beltrami operator associated with the Riemannian metric. 
The definition allows for an interpretation of the mean as an extension of the Fréchet mean due to Varadhan's result stating that $\lim_{t \to 0}-2 t \log p_t(x,y) =  d(x,y)^2$ uniformly on compact sets disjoint from the cut-locus of either $x$ or $y$ \cite{hsu2002stochastic}.

Just as the Fr\'echet mean, the diffusion mean has a weighted version, and the corresponding estimator of the weighted diffusion $t$-mean is given as
    \begin{equation}
        \hat \mu_t =  \argmin_{p \in M}  \sum_{i=1}^n - \log p_{t/w_i}(p,x_i).
        \label{eq:wdiffusionmean}
    \end{equation}
    Note that the evaluation time is here scaled by the weights. This is equivalent to scaling the variance of the steps of the Brownian motion \cite{grong_most_2021}.
     
As closed-form expressions for the heat kernel are only available on specific manifolds, evaluating the diffusion $t$-mean often rely on numerical methods. 
One example of this is using bridge sampling to numerically estimate the transition density \cite{sommer_bridge_2017,pennec_riemannian_2020}. If a global coordinate chart is available, the transition density can be written in the form (see \cite{papaspiliopoulos2012importance, jensen2021simulation})
    \begin{equation}\label{eq: transition density estimate}
        p_T(z,v) = \sqrt{\frac{\det g(v)}{(2\pi T)^2}}e^{- \frac{\lVert a(z)(z-v)\rVert^2}{2T}} \E\left[\varphi\right],
    \end{equation}
where $g$ is the metric matrix, $a$ a square root of $g$, and $\varphi$ denotes the correction factor between the law of the true diffusion bridge and the law of the simulation scheme. The expectation over the correction factor can be numerically approximated using Monte Carlo sampling. The correction factor will appear again when we discuss guided bridge proposals below.

\subsection{Diffusion bridges}
The proposed sampling scheme for the (weighted) diffusion mean builds on simulation methods for conditioned diffusion processes, diffusion bridges. We here outline ways to simulate conditioned diffusion processes numerically in both the Euclidean and manifold context.

\subsubsection{Euclidean diffusion bridges}
Let $(\Omega, \cF, \cF_t, \Pb)$ be a filtered probability space, and $X$ a $d$-dimensional Euclidean diffusion $[0,T]$ satisfying the stochastic differential equation (SDE) 
    \begin{equation}
        dX_t = b_t(X_t)dt + \sigma_t(X_t) dW_t, \quad X_0 = x,
    \end{equation}
where $W$ is a $d$-dimensional Brownian motion. Let $v\in\mathbb R^d$ be a fixed point. Conditioning $X$ on reaching $v$ at a fixed time $T>0$ gives the bridge process $X|X_T=v$. Denoting this process $Y$, Doob's $h$-transform shows that $Y$ is a solution of the SDE (see e.g. \cite{lyons_conditional_1990})
    \begin{equation}
        \begin{split}
        dY_t &=  \tilde b_t(Y_t)dt + \sigma_t(Y_t)d\tilde W_t , \quad Y_0 = x\\
        \tilde b_t(y) &= b_t(y) +  a_t(y) \nabla_y \log p_{T-t}(y,v), 
        \end{split}\label{eq:doob}
    \end{equation}
where $p_t(\cdot, \cdot)$ denotes the transition density of the diffusion $X$, $a = \sigma \sigma^T$, and where $\tilde W$ is a $d$-dimensional Brownian motion under a changed probability law. 
From a numerical viewpoint, in most cases, the transition density is intractable and therefore direct simulation of \eqref{eq:doob} is not possible. 

If we instead consider a Girsanov transformation of measures to obtain the system (see, e.g., \cite[Theorem 1]{delyon_simulation_2006})
    \begin{equation}
    \begin{split}
        dY_t &=  \tilde b_t(Y_t)dt + \sigma_t(Y_t)d\tilde W_t , \quad Y_0 = x\\
        \tilde b_t(y) &= b_t(y) +  \sigma_t(y) h(t,y), 
    \end{split}
    \label{eq: girsanov's sde transform 2}
    \end{equation}
the corresponding change of measure is given by
    \begin{equation}
        \frac{d\Pb^h}{d\Pb}\bigg|_{\cF_t} = e^{\int_0^t h(s,X_s)^T dW_s - \frac{1}{2}\int_0^t \lVert h(s,X_s)\rVert^2 ds}.
    \end{equation}
From \eqref{eq:doob}, it is evident that $h(t,x) = \sigma^T \nabla_x \log p_{T-t}(x,v)$ gives the diffusion bridge. However, different choices of the function $h$ can yield processes which are absolutely continuous wrt. to the actual bridges, but which can be simulated directly. 

Delyon and Hu \cite{delyon_simulation_2006} suggested to use $h(t,x)= \sigma_t^{-1}(x) \nabla_x \log q_{T-t}(x,v)$, where $q$ denotes the transition density of a standard Brownian motion with mean $v$, i.e., $q_t(x,v) = (2\pi t)^{-d/2}\exp(-\lVert x-v \rVert^2/2t)$. They furthermore proposed a method that would disregard the drift term $b$, i.e.,  $h(t,x) )= \sigma_t^{-1}(x) \nabla_x \log q_{T-t}(x,v) - \sigma_t^{-1}(x)b_t(x)$.
Under certain regularity assumptions on $b$ and $\sigma$, the resulting processes converge to the target in the sense that $\lim_{t \to T} Y_t = v$ a.s. In addition, for bounded continuous functions $f$, the conditional expectation is given by
    \begin{equation}
        \E\left[f(X)|X_T=v\right] = C \E\left[f(Y)\varphi(y) \right],
    \end{equation}
where $\varphi$ is a functional of the whole path $Y$ on $[0,T]$ that can be computed directly. From the construction of the $h$-function, it can be seen that the missing drift term is accounted for in the correction factor $\varphi$.

The simulation approach of \cite{delyon_simulation_2006} can be improved by the simulation scheme introduced by Schauer et al. \cite{schauer2017guided}. Here, an $h$-function defined by $h(t,x) = \nabla_x \log \hat p_{T-t}(x,v)$ is suggested, where $\hat p$ denotes the transition density of an auxiliary diffusion process with known transition densities. The auxiliary process can for example be linear because closed-form solutions of transition densities for linear processes are available. Under the appropriate measure $\Pb^h$, the guided proposal process is a solution to 
    \begin{equation}
        dY_t = b_t(Y_t)dt + a_t(Y_t) \nabla_x \log \hat p_{T-t}(x,v)|_{x=Y_t}dt + \sigma_t(Y_t)dW_t.
        \label{eq:schauer}
    \end{equation}
Note the factor $a(t,y)$ in the drift in \eqref{eq:doob} which is also present in \eqref{eq:schauer} but not with the scheme proposed by \cite{delyon_simulation_2006}. Moreover, the choice of a linear process grants freedom to model.
For other choices of an $h$-functions see e.g. \cite{marchand2011conditioning,van2017bayesian}.

Marchand \cite{marchand2011conditioning} extended the ideas of Delyon and Hu by conditioning a diffusion process on partial observations at a finite collection of deterministic times. Where Delyon and Hu considered the guided diffusion processes satisfying the SDE
    \begin{equation}\label{eq: delyon hu sde}
        dY_t = b_t(Y_t)dt - \frac{Y_t-v}{T-t}dt + \sigma_t(Y_t)dw_t,
    \end{equation}
for $v \in \R^d$ over the time interval $[0,T]$, Marchand proposed the guided diffusion process conditioned on partial observations ${v_1,\dots, v_N}$ solving the SDE
    \begin{equation}\label{eq: marchand sde}
        dY_t = b_t(Y_t)dt - \sum_{k=1}^n P_t^k(Y_t)\frac{Y_t-u_k}{T_k-t}1_{(T_k - \varepsilon_k,T_k)}dt + \sigma_t(Y_t)dw_t,
    \end{equation}
    where $u_k$ is be any vector satisfying $L_k(x)u_k = v_k$, $L_k$ a deterministic matrix in $M_{m_k,n}(\R)$ whose $m_k$ rows form a orthonormal family, $P^k_t$ are projections to the range of $L_k$, and $T_k-\varepsilon_k < T_k$. The $\varepsilon_k$ allow to only apply the guiding term on a part of the time intervals $[T_{k-1},T_k]$. We will only consider the case $k=1$. The scheme allows to sample bridges conditioned on $LY_T = v$.

\subsection{Manifold diffusion processes}
To work with diffusion bridges and guided proposals on manifolds, we will first need to consider the Eells-Elworthy-Malliavin construction of Brownian motion and the connected characterization of semimartingales \cite{elworthy1988geometric}. Endowing the frame bundle $FM$ with a connection allows splitting the tangent bundle $TFM$ into a horizontal and vertical part. If the connection on $FM$ is a lift of a connection on $M$, e.g. the Levi-Civita connection of a metric on $M$, the horizontal part of the frame bundle is in one-to-one correspondence with $M$. In addition, there exist fundamental horizontal vector fields $H_i \colon  FM \rightarrow HFM$ such that for any continuous $\R^d$-valued semimartingale $Z$ the process $U$ defined by
    \begin{equation}
        dU_t = H_i(U_t) \circ dZ^i_t,
        \label{eq:FMsemi}
    \end{equation}
is a horizontal frame bundle semimartingale, where $\circ$ denotes integration in the Stratonovich sense. The process $X_t := \pi(U_t)$ is then a semimartingale on $M$. Any semimartingale $X_t$ on $M$ has this relation to a Euclidean semimartingale $Z_t$. $X_t$ is denoted the development of $Z_t$, and $Z_t$ the antidevelopment of $X_t$. We will use this relation when working with bridges on manifolds below.

When $Z_t$ is a Euclidean Brownian motion, the development $X_t$ is a Brownian motion. We can in this case also consider coordinate representations of the process. With an atlas $\{(D_{\alpha}, \phi_{\alpha})\}_{\alpha}$ of $M$, there exists an increasing sequence of predictable stopping times $0 \leq T_k \leq T_{k+1}$ such that on each stochastic interval $\llbracket T_k, T_{k+1}\rrbracket = \{(\omega,t)\in \Omega \times \R_+| T_k(\omega) \leq t \leq T_{k+1}(\omega)\}$ the process $x_t \in D_{\alpha}$, for some $\alpha$ (see \cite[Lemma 3.5]{emery1989stochastic}). Thus, the Brownian motion $x$ on $M$ can be described locally in a chart $D_{\alpha} \subset M$ as the solution to the system of SDEs, for $(\omega,t) \in \llbracket T_k, T_{k+1} \rrbracket \cap \{T_k < T_{k+1}\}$
    \begin{equation}\label{eq: local brownian motion}
        dx_t^i(\omega) = b^i(x_t(\omega))dt + \sigma^i_j(x_t(\omega)) dW^j_t(\omega), 
    \end{equation}
where $\sigma$ denotes the matrix square root of the inverse of the Riemannian metric tensor $(g^{ij})$ and $b^k(x) = - \frac{1}{2}g^{ij}(x)\Gamma^k_{ij}(x)$ is the contraction over the Christoffel symbols (see, e.g., \cite[Chapter 3]{hsu2002stochastic}). Strictly speaking, the solution of equation \eqref{eq: local brownian motion} is defined by $x_t^i = \phi_{\alpha}(x_t)^i$. 

We thus have two concrete SDEs for the Brownian motion in play: The $FM$ SDE \eqref{eq:FMsemi} and the coordinate SDE \eqref{eq: local brownian motion}.

Throughout the paper, we assume that $M$ is stochastically complete, i.e. the Brownian motions does not explode in finite time and, as a consequence, $\int_M p_t(x,y) d\Vol_M(y) = 1$, for all $t>0$ and all $x \in M$.

\subsection{Manifold bridges}\label{sec: bridges on manifolds}
The Brownian bridge process $Y$ on $M$ conditioned at $Y_T = v$ is a Markov process with generator $\frac{1}{2}\Delta + \nabla \log p_{T-t}(\cdot, v)$. Closed-form expressions of the transition density of a Brownian motion are available on selected manifolds including Euclidean spaces, hyperbolic spaces, and hyperspheres. Direct simulation of Brownian bridges is therefore possible in these cases. However, generally, transition densities are intractable and auxiliary processes are needed to sample from the desired conditional distributions.  

To this extent, various types of bridge processes on Riemannian manifolds have been described in the literature. In case of manifolds with a pole, i.e, the existence of a point $p \in M$ such that the exponential map $\exp_p \colon T_pM \rightarrow M$ is a diffeomorphism, the \textit{semi-classical} (Riemannian Brownian) bridge was introduced by Elworthy and Truman \cite{elworthy1982diffusion} as the process with generator $\frac{1}{2}\Delta + \nabla \log k_{T-t}(\cdot, v)$, where
    \begin{equation*}
        k_t(x,v) = (2\pi t)^{-n/2} e^{-\frac{d(x,v)^2}{2t}} J^{-1/2}(x),
    \end{equation*}
and $J(x) = |\det D_{\exp(v)^{-1}} \exp_{v}|$ denotes the Jacobian determinant of the exponential map at $v$. Elworthy and Truman used the semi-classical bridge to obtain heat kernel estimates, and the semi-classical bridge has been studied by various others \cite{li2017semi,ndumu1991brownian}. 

By Varadhan's result (see \cite[Theorem 5.2.1]{hsu2002stochastic}), as $t \rightarrow T$, we have the asymptotic relation $((T-t) \log p_{T-t}(x,y) \sim  -\frac{1}{2}d(x,y)^2$. In particular, the following asymptotic relation was shown to hold by Malliavin, Stroock, and Turetsky \cite{malliavin1996short,stroock1997short} : $(T-t)\nabla \log p_{T-t}(x,y) \sim - \frac{1}{2} \nabla d(x,y)^2$. From these results, the generators of the Brownian bridge and the semi-classical bridge differ in the limit by a factor of $-\frac{1}{2}\nabla \log J(x)$. However, under a certain boundedness condition, the two processes can be shown to be identical under a changed probability measure \cite[Theorem 4.3.1]{thompson2015submanifold}.

In order to generalize the heat-kernel estimates of Elworthy and Truman, Thompson \cite{thompson2015submanifold,thompson2018brownian} considered the \textit{Fermi} bridge process conditioned to arrive in a submanifold $N \subseteq M$ at time $T>0$. The Fermi bridge is defined as the diffusion process with generator $\frac{1}{2}\Delta + \nabla \log q_{T-t}(\cdot, N)$, where
    \begin{equation*}
        q_t(x,N) = (2\pi t)^{-n/2} e^{-\frac{d(x,N)^2}{2t}}.
    \end{equation*}
For both of these bridge processes, when $M = \R^d$ and $N$ is a point, both the semi-classical bridge and the Fermi bridge agree with the Euclidean Brownian bridge.

\cite{jensen2021simulation} introduce a numerical simulation scheme for conditioned diffusions on Riemannian manifolds, which generalize the method by Delyon and Hu \cite{delyon_simulation_2006}. The guiding term used is identical to the guiding term of the Fermi bridge when the submanifold is a single point $v$.

\section{Diffusion mean estimation}
The standard setup for diffusion mean estimation described in the literature (e.g. \cite{sommer_bridge_2017}) is as follows: Given a set of observations $x_1,\dots,x_n \in M$, for each observation $x_i$, sample a guided bridge process approximating the bridge $X_{i,t}|X_{i,T} = x_i$ with starting point $x_0$. The expectation over the correction factors can be computed from the samples, and the transition density can be evaluated using \eqref{eq: transition density estimate}. An iterative maximum likelihood approach using gradient descent to update $x_0$ yielding an approximation of the diffusion mean in the final value of $x_0$. The computation of the diffusion mean, in the sense just described, is, similarly to the Fréchet mean, computationally expensive. 

We here explore the idea first put forth in \cite{sommer2020horizontal}: We turn the situation around to simulate $n$ independent Brownian motions starting at each of $x_1,\ldots,x_n$, and we condition the $n$ processes to coincide at time $T$. We will show that the value $x_{1,T}=\cdots=x_{n,T}$ is an estimator of the diffusion mean. By introducing weights in the conditioning, we can similarly estimate the weighted diffusion mean. The construction can concisely be described as a single Brownian motion on the $n$-times product manifold $M^n$ conditioned to hit the diagonal $\diag(M^n)=\{(x,\ldots,x)|x\in M\}\subset M^n$. 
To shorten notation, we denote the diagonal submanifold $N$ below.
We start with examples with $M$ Euclidean to motivate the construction.

\begin{Example}
    Consider the two-dimensional Euclidean multivariate normal distribution 
    \begin{equation*}
    \begin{pmatrix}
        X\\Y
        \end{pmatrix}
        \sim
        N\left(
        \begin{pmatrix}
        \mu_1\\ \mu_2
        \end{pmatrix}
        ,
        \begin{pmatrix}
        \sigma_{11} \quad \sigma_{12}\\\sigma_{21} \quad \sigma_{22}
        \end{pmatrix}
        \right).
    \end{equation*}
The conditional distribution of $X$ given $Y=y$ follows a univariate normal distribution
    \begin{equation*}
        X|Y=y \sim N\left(\mu_1 + \sigma_{12}\sigma_{22}^{-1}(y-\mu_2), \sigma_{11}-\sigma_{12}\sigma_{22}^{-1}\sigma_{21}\right).
    \end{equation*}
This can be seen from the fact that if $X \sim N\left(\mu, \Sigma\right)$ then for any linear transformation $AX + b \sim N\left(b + A\mu, A \Sigma A^T\right)$. Defining the random variable $Z= X-\sigma_{12}\sigma_{22}^{-1}Y$, the result applied to $(Z,X)$ gives $Z \sim N\left(\mu_1 - \sigma_{12}\sigma_{22}^{-1}\mu_2, \sigma_{11}-\sigma_{12}\sigma_{22}^{-1}\sigma_{21}\right)$. The conclusion then follows from $X= Z +\sigma_{12}\sigma_{22}^{-1}Y$. Note that $X$ and $Y$ are independent if and only if $\sigma_{12}=\sigma_{21}=0$ and the conditioned random variable is in this case identical in law to $X$.
\end{Example}

Let now $x_1, \dots, x_n \in M$ be observations and let $x = (x_1, \dots, x_n) \in M^n$ be an element of the $n$-product manifold $M \times \dots \times M$ with the product Riemannian metric. We again first consider the case $M=\mathbb R^d$:

\begin{Example}
  Let $Y_i \sim N\left(x_i, \frac{T}{w_i} I_d\right)$ be independent random variables. The conditional distribution $Y_1|Y_1= \dots =Y_n$ is normal $N\left( \frac{\sum_i w_i x_i}{\sum_i w_i},\frac{T}{\sum_i w_i}\right)$. This can be seen inductively:   
    The conditioned random variable $Y_1 | Y_1 = Y_2$ is identical to $Y_1 | Y_1-Y_2 = 0$. Now let $X = Y_1$ and $Y = Y_1-Y_2$ and refer to Example 1.
    In order to conclude, assume $Z_n := Y_1|Y_1 = \cdots, = Y_{n-1}$ follows the desired normal distribution. Then $Z_n|Z_n = Y_n$ is normally distributed with the desired parameters and $Z_n|Z_n = Y_n$ is identical to $Y_1|Y_1=\cdots = Y_n$.
\end{Example}

The following example establishes the weighted average as a projection onto the diagonal.

\begin{figure}
\centering
\def\svgwidth{0.3\textwidth}
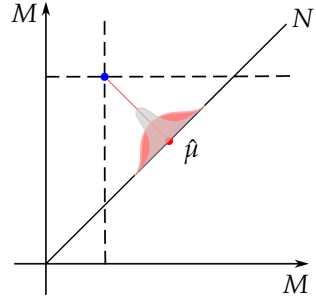
\caption{The mean estimator viewed as a projection onto the diagonal of a product manifold. Conditioning on the closest point in the diagonal yields a density on the diagonal depending on the time to arrival $T>0$. As $T$ tends to zero the density convergence to the Dirac-delta distribution (grey), whereas as $T$ increases the variance of the distribution increases (rouge).}\label{fig: projection on diagonal with density}
\end{figure}

\begin{Example}
    Let $x$ be a point in $(\R^d)^n$ and let $P$ be the orthogonal projection to the diagonal of $(\R^d)^n$ such that $Px = \left(\frac{1}{nd} \sum_{i=1}^{nd} x_i \dots \frac{1}{nd} \sum_{i=1}^{nd} x_i\right)^T$. We see that the projection yields $n$ copies of the arithmetic mean of the coordinates. This is illustrated in Figure~\ref{fig: projection on diagonal with density}.
\end{Example}

The idea of conditioning diffusion bridge processes on the diagonal of a product manifold originates from the facts established in examples 1-3.
We sample the mean by sampling from the conditional distribution $Y_1|Y_1=\dots=Y_n$ from example 2 using a guided proposal scheme on the product manifolds $M^n$ and on each step of the sampling projecting to the diagonal as in example 3. 

Turning now to the manifold situation, we replace the normal distributions with mean $x_i\in\R^d$ and variance $T/w_i$ with Brownian motions started at $x_i\in M$ and evaluated at time $T/w_i$. Note that the Brownian motion density, the heat kernel, is symmetric in its coordinates: $p_t(x,y)=p_t(y,x)$. We will work with multiple process and indicate with superscript the density with respect to a particular process, e.g. $p_T^{X}$. Note also that change of the evaluation time $T$ is equal to scaling the variance, i.e. $p_{\alpha T}^X(x,y)=p_T^{X^\alpha}(x,y)$ where $X^\alpha$ is a Brownian motion with variance of the increments scaled by $\alpha>0$. This gives the following theorem, first stated in \cite{sommer2020horizontal} with sketch proof:
\begin{Theorem}
  Let $X_t=(X_{1,t}^{w_1^{-1}},\dots,X_{n,t}^{w_n^{-1}})$ consist of $n$ independent Brownian motions on $M$ with variance $w_i^{-1}$ and $X_{i,0}=x_i$, and let $\mathbb P^*$ the law of the conditioned process $Y_t=X_t|X_T\in N$, $N=\diag(M^n)$. Let $v$ be the random variable $Y_{1,T}$. Then $v$ has density $p_v^Y(y)\propto \prod_{i=1}^np_{T/w_i}(x_i; y)$ and $v=Y_{i,T}$ for all $i$ a.s. (almost surely).

  \label{thm:product_processes}
\end{Theorem}
\begin{proof}
  $p_T^X((x_1,\dots,x_n),(y,\dots,y))=\prod_{i=1}^np_T^{X^{w_i^{-1}}}(x_i, y)$
  because the processes $X_{i,t}$ are independent. By symmetry of the Brownian motion and the time rescaling property, $p_T^{X_i^{w_i^{-1}}}(x_i, y)=p_{T/w^i}(y, x_i)$. For elements $(y,\dots,y)\in\diag(M^n)$ and $x\in M^n$, $p_v(y)=p_T^Y(x,y)\propto p_T^X(x,y)$. As a result of the conditioning, $v=Y_{1,T}=\cdots=Y_{n,T}$. In combination, this establishes the result.
\end{proof}
As a consequence, the set of modes of $p_v$ equal the set of the maximizers for the likelihood $L(y; x_1,\dots,x_n)=\prod_{i=1}^np_{T/w_i}(x_i; y)$ and thus the weighted diffusion mean. This result is the basis for the sampling scheme. Intuitively, if the distribution of $v$ is relatively well behaved (e.g. close to normal), a sample from $v$ will be close to a weighted diffusion mean with high probability.

In practice, however, we cannot sample $Y_t$ directly. Instead, we will below use guided proposal schemes resulting in processes $\tilde{Y}_t$ with law $\tilde{\mathbb P}$ that we can actually sample and that, under certain assumptions, will be absolutely continuous with respect to $Y_t$ with explicitly computable likelihood ratio so that $\mathbb P^*=\tfrac{\varphi(\tilde Y_T)}{\E^{\tilde{\mathbb P}}[\varphi(\tilde Y_T)]} \tilde{\mathbb P}$.
\begin{Corollary}
  Let $\tilde{\mathbb P}$ be the law of $\tilde{Y}_t$ and $\varphi$ be the corresponding correction factor of the guiding scheme. Let $\tilde{v}$ be the random variable $\tilde{Y}_{1,T}$ with law $\tfrac{\varphi(\tilde Y_T)}{\E^{\tilde{\mathbb P}}[\varphi(\tilde Y_T)]} \tilde{\mathbb P}$. Then $\tilde{v}$ has density $p_{\tilde{v}}(y)\propto \prod_{i=1}^np_{T/w_i}(x_i; y)$.
\end{Corollary}
We now proceed to actually construct the guided sampling schemes.

\subsection{Fermi bridges to the diagonal}
Consider a Brownian motion $X_t=(X_{1,t}, \dots X_{n,t})$ in the product manifold $M^n$ conditioned on $X_{1,T} = \dots = X_{n,T}$ or, equivalently, $X_T\in N$, $N=\diag(M^n)$. 
Since $N$ is a submanifold of $M^n$, the conditioned diffusion defined above is absolutely continuous with respect to the Fermi bridge on $[0,T)$ \cite{thompson2015submanifold, thompson2018brownian}. Define the $FM$-valued horizontal guided process
    \begin{equation}\label{eq: fermi bridge}
        dU_t = H_i(U_t) \circ \left(dW^i_t - \frac{H_i \tilde r_N^2(U_t) }{2(T-t)}dt \right),
    \end{equation}
where $\tilde r$ denotes the lift of the radial distance to $N$ defined by $\tilde r_N(u) := r_N(\pi(u))= d(\pi(u), N)$. 
The Fermi bridge $Y^F$ is the projection of $U$ to $M$, i.e., $Y_t^F := \pi(U_t)$. Let $\Pb^F$ denotes its law.

\begin{Theorem}
    For all continuous bounded functions $f$ on $M^n$, we have
        \begin{equation}
          \E\left[f(X)|X_{1,T} = \cdots = X_{n,T}\right] =  \lim_{t \uparrow T}C\E^{\Pb^F}\left[f(Y^F)\varphi(Y^F_t)\right], 
            \label{eq:diagthm}
        \end{equation}
    with a constant $C>0$, where 
       \begin{equation*}
			d\log \varphi(Y^F_s) = 
	\frac{r_N(Y_s^F)}{T-s} \left(d\eta_s + dL_s\right) \quad \text{with} \quad 	d\eta_s = \frac{\partial}{\partial r_N} \log \Theta_N^{-\frac{1}{2}}(Y^F_s)ds,
		\end{equation*}
        $dL_s := d\mathbb L_s(Y^F)$ with $\mathbb L$ being the geometric local time at $\Cut(N)$, and $\Theta_N$ is the determinant of the derivative of the exponential map normal to $N$ with support on $M^n\backslash \Cut(N)$ \cite{thompson2015submanifold}.
    \label{thm:diag}
\end{Theorem}

\begin{proof}
From \cite[Theorem 8]{jensen2021simulation} and \cite{thompson2018brownian}, 
    \begin{equation*}
      \E\left[f(X)|X_T \in N\right] = \lim_{t \uparrow T}C\E^{\Pb^F}\left[f(Y^F)\varphi(Y_t^F)\right]
      .
    \end{equation*}
\end{proof}
Since $N$ is a totally geodesic submanifold of dimension $d$, the results of \cite{thompson2015submanifold} can be used to give sufficient conditions to extend the equivalence in \eqref{eq:diagthm} to the entire interval $[0,T]$. A set $A$ is said to be polar for a process $X_t$ if the first hitting time of $A$ by $X$ is infinity a.s.

\begin{Corollary}\label{result: james corollary}
    If either of the following conditions are satisfied
    \begin{itemize}
        \item[i)] the sectional curvature of planes containing the radial direction is non-negative \textit{or} the Ricci curvature in the radial direction is non-negative;
        \item[ii)] $\Cut(N)$ is polar for the Fermi bridge $Y^F$ and either the sectional curvature of planes containing the radial direction is non-positive or the Ricci curvature in the radial direction is non-positive;
    \end{itemize}
then
    \begin{equation*}
      \E\left[f(X)|X_{1,T} = \cdots = X_{n,T}\right] =  C\E^{\Pb^F}\left[f(Y^F)\varphi(Y_T^F)\right].
    \end{equation*}
    In particular, $\frac{\varphi(Y^F_T)}{\E^{\Pb^F}[\varphi(Y^F_T)]} d\Pb^F \propto d\Pb^*$.
\end{Corollary} 
\begin{proof}
See \cite[Appendix C.2]{thompson2015submanifold}.
\end{proof}

For numerical purposes, the equivalence \eqref{eq:diagthm} in Theorem~\ref{thm:diag} is sufficient as the interval $[0,T]$ is finitely discretized. To get the result on the full interval, the conditions in \refcor{result: james corollary} may at first seem quite restrictive. A sufficient condition for a subset of a manifold to be polar for a Brownian motion is its Hausdorff dimension being two less than the dimension of the manifold. Thus, $\Cut(N)$ is polar if $\dim(\Cut(N))\le nd-2$. Verifying whether this is true requires specific investigation of the geometry of $M^n$.

The SDE \eqref{eq: fermi bridge} together with \eqref{eq:diagthm} and the correction $\varphi$ gives a concrete simulation scheme that can be implemented numerically. Implementation of the geometric constructs is discussed in section \ref{sec:experiments}. The main complication of using Fermi bridges for simulation is that it involves evaluation of the radial distance $r_N$ at each time-step of the integration. Since the radial distance finds the closest point on $N$ to $x_1,\dots,x_n$, it is essentially a computation of the Fr\'echet mean and thus hardly more computationally efficient than computing the Fr\'echet mean itself. For this reason, we present a coordinate based simulation scheme below.

\subsection{Simulation in coordinates}
We here develop a more efficient simulation scheme focusing on manifolds that can be covered by a single chart. The scheme follows the partial observation scheme developed  \cite{marchand2011conditioning}. Representing the product process in coordinates and using a transformation $L$, whose kernel is the diagonal $\diag(M^n)$, gives a guided bridge process converging to the diagonal. An explicit expression for the likelihood is given. 
    
In the following, we assume that $M$ can be covered by a chart in which the square root of the cometric tensor, denoted by $\sigma$, is $C^2$. Furthermore, $\sigma$ and its derivatives are bounded; $\sigma$ is invertible with bounded inverse. The drift $b$ is locally Lipschitz and locally bounded.

Let $x_1, \dots, x_n \in M$ be observations and let $X_{1,t}, \dots, X_{n,t}$ be independent Brownian motions with $X_{1,0}=x_1,\dots, X_{n,0}=x_n$. Using the coordinate SDE \eqref{eq: local brownian motion} for each $X_{i,t}$, we can write the entire system on $M^n$ as
\begin{equation}\label{eq: local brownian motion on product manifold}
    d
    \begin{pmatrix}
        X^1_{1,t} \\ 
        \vdots \\ 
        X^d_{1,t}\\
        \vdots \\
        X^1_{n,t} \\
        \vdots \\
        X^d_{n,t}
    \end{pmatrix}
        =
        \begin{pmatrix}
            b^1(X_{1,t}) \\ 
            \vdots \\ 
            b^d(X_{1,t})\\
            \vdots \\
            b^1(X_{n,t}) \\ 
            \vdots \\
            b^d(X_{n,t})
        \end{pmatrix}
        dt
        +
        \left[\begin{array}{ccc}
\Sigma^{1}(X_{1,t},\dots,X_{n,t}) \\
\vdots   \\
\Sigma^d(X_{1,t},\dots,X_{n,t}) \\
\end{array}\right] d W_t.
\end{equation}
In the product chart, $\Sigma$ and $b$ satisfy the same assumptions as the metric and cometric tensor and drift listed above.

The conditioning $X_T\in N$ is equivalent to the requiring $X_T\in\diag((\mathbb R^d)^n)$ in coordinates. $\diag((\mathbb R^d)^n)$ is a linear subspace of $(\mathbb R^d)^n$, we let
$L\in M^{d\times nd}$ be a matrix with orthonormal rows and $\ker L = \diag((\mathbb R^d)^n)$ so that the desired conditioning reads $LX_T = 0$. Define the following oblique projection, similar to \cite{marchand2011conditioning},
\begin{equation}
    P_t(x) = a(x)L^T A(x) L
\end{equation}
where
    \begin{equation*}
        a(x) = \Sigma(x)\Sigma(x)^T \quad \text{and} \quad A_t(x) = (La(x)L^T)^{-1}.
    \end{equation*}
Set $\beta(x) = \Sigma(x)^T L^T A(x)$. The guiding scheme \eqref{eq: marchand sde} then becomes
    \begin{equation}\label{eq: marchand SDE on manifold}
        dY_t = b(Y_t)dt + \Sigma(Y_t)d W_t - \Sigma(Y_t)\beta(Y_t)  \frac{LY_t}{T-t} 1_{(T-\varepsilon,T)}(t)dt, \quad Y_0 = u.
    \end{equation}
We have the following result.    
    
\begin{Lemma}\label{result: convergence lemma marchand}
Equation \eqref{eq: marchand SDE on manifold} admits a unique solution on $[0,T)$. Moreover, $\lVert L Y_t \rVert^2 \leq C(\omega) (T-t) \log \log [(T-t)^{-1} + e]$ a.s., where $C$ is a positive random variable.
\end{Lemma}

\begin{proof}
    Since $L P = L$, the proof is similar to the proof of  \cite[Lemma 6]{marchand2011conditioning}.
\end{proof}

With the same assumptions, we get as well the following result similar to \cite[Theorem 3]{marchand2011conditioning}.
\begin{Theorem}
 Let $Y_t$ be a solution of \eqref{eq: marchand SDE on manifold}, and assume the drift $b$ is bounded. For any bounded function $f$,
    \begin{equation}
        \E\left[f(X)|X_T\in N\right] 
        =
        C\E\left[f(y) \varphi(Y)\right], 
    \end{equation}
where $C$ is a positive constant and 
    \begin{equation*}
      \begin{split}
        \varphi(Y_t) 
        &
        =
        \sqrt{\det(A(Y_T))} \exp\bigg\{- \frac{\lVert \beta_{T-\varepsilon}(Y_{T-\varepsilon})LY_{T-\varepsilon}\rVert^2}{2\varepsilon} 
        \\
         -
         &
         \int_{T-\varepsilon}^T \frac{2(LY_s)^TL b(Y_s)ds - (LY_s)^Td(A(Y_s)) LY_s + d[A(Y_s)^{ij}, (LY_s)_i(LY_s)_j]}{2(T-s)}
        \bigg\}
      \end{split}
    \end{equation*}
\end{Theorem}

\begin{proof}
    A direct consequence of \cite[Theorem 3]{marchand2011conditioning}, for $k=1$, and \reflemma{result: convergence lemma marchand}.
\end{proof}
The theorem can also be applied for unbounded drift by replacing $b$ with a bounded approximation and performing a Girsanov change of measure.

\begin{algorithm} \DontPrintSemicolon \SetAlgoLined
\textbf{Input:} Points $x_1,\dots, x_n \in M$
\textbf{Output:} (weighted) diffusion mean sampling\\
\For{$j=1$ to $J$}{
Sample path from guided process $Y_t$\\
Record $Y_T^j$ and compute correction factor $\varphi(Y^j_T)$
}
Sample $j$ from $1,\dots, J$ with probability $P_j = \frac{\varphi(Y^j_T)}{\sum_{k=1}^J \varphi(Y^k_T)}$.\\
\tcp{Return $Y_T^j$}
\caption{weighted Diffusion Mean}
\label{alg: weighted diffusion mean}
\end{algorithm}
    
\subsection{Accounting for $\varphi$}
The sampling schemes \eqref{eq: fermi bridge}, \eqref{eq: marchand SDE on manifold} above provides samples on the diagonal and thus candidates for the diffusion mean estimates. However, the schemes sample from a distribution which is only equivalent to the bridge process distribution: We still need to handle the correction factor in the sampling to sample from the correct distribution, i.e. the rescaling $\tfrac{\varphi}{\E[\varphi]}$ of the guided law in Theorem~\ref{thm:product_processes}. A simple way to achieve this is to do sampling importance resampling (SIR) as described in Algorithm~\ref{alg: weighted diffusion mean}. This yields an approximation of the weighted diffusion mean. For each sample $y^i$ of the guided bridge process, we compute the corresponding correction factor $\varphi(y^i)$. The resampling step then consists in picking $y_T^j$ with a probability determined by the correction terms, i.e., with $J$ the number of samples we pick sample $j$ with probability $P_j = \frac{\varphi(y^j_T)}{\sum_{i=1}^J \varphi(y^i_T)}$. 

It depends on the practical application if the resampling is necessary, or if direct samples from the guided process (corresponding to $J=1$) are sufficient.

\section{Experiments}
\label{sec:experiments}
We here exemplify the mean sampling scheme on the two-sphere $\mathbb S^2$ and on finite sets of landmark configurations endowed with the LDDMM metric \cite{joshi_landmark_2000,younes_shapes_2010}. With the experiment on $\mathbb S^2$, we aim to give a visual intuition of the sampling scheme and the variation in the diffusion mean estimates caused by the sampling approach. In the higher-dimensional landmark example where closed-form solutions of geodesics are not available, we compare to the Fr\'echet mean and include rough running times of the algorithms to give a sense of the reduced time complexity. Note, however, that the actual running times are very dependent on the details of the numerical implementation, stopping criteria for the optimization algorithm for the Fr\'echet mean, etc.
\begin{figure}[t]
    \centering
    \includegraphics[width=0.5\linewidth,clip=true,trim=100 200 100 200]{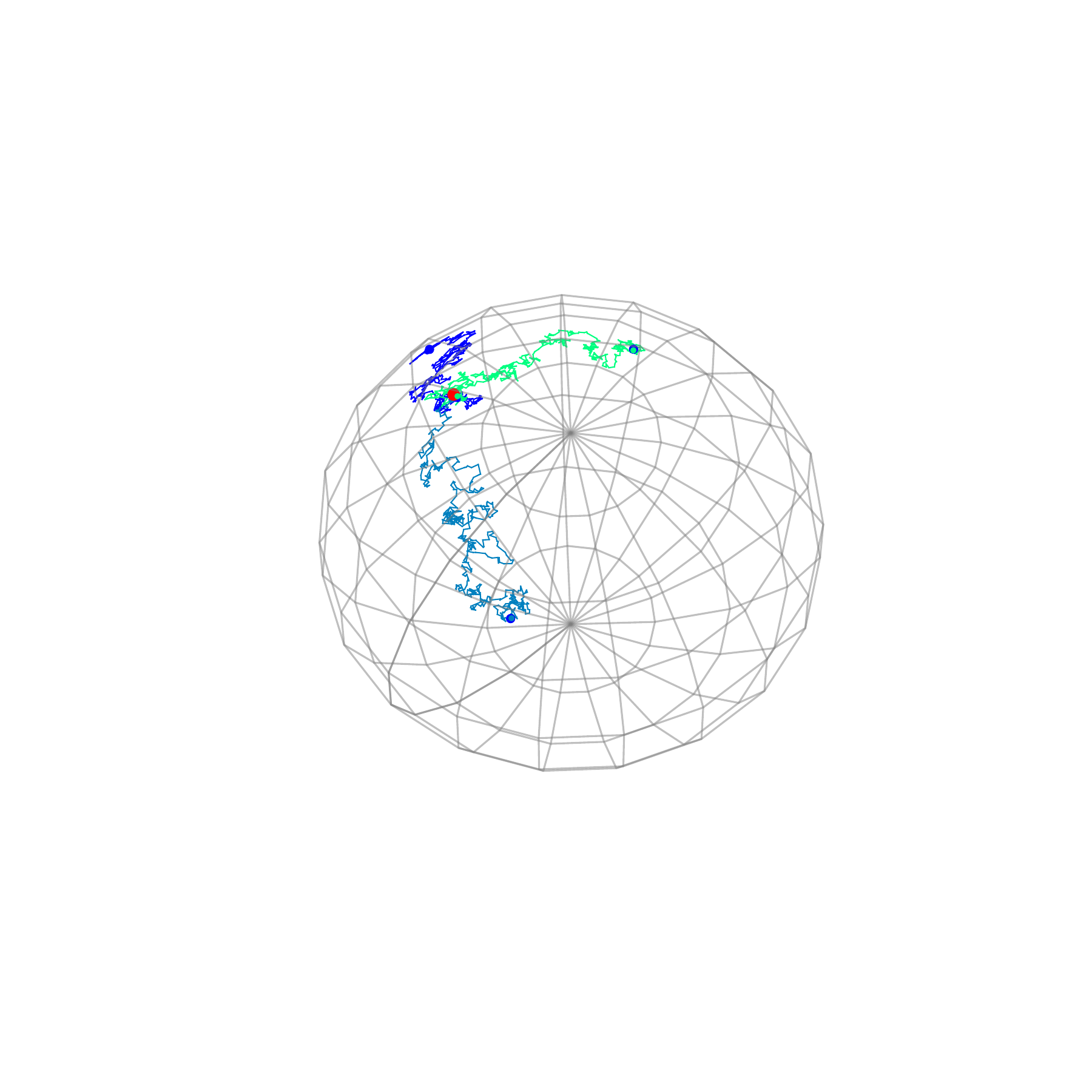}
    \caption{3 points on $\mathbb S^2$ together with a sample mean (red) and the diagonal process in $(\mathbb S^2)^n$, $n=3$ with $T=.2$ conditioned on the diagonal.}\label{fig:s2-1}
\end{figure}
\begin{figure}[t]
  \subfloat{
        \includegraphics[width=.48\linewidth,clip=true,trim=100 150 100 200]{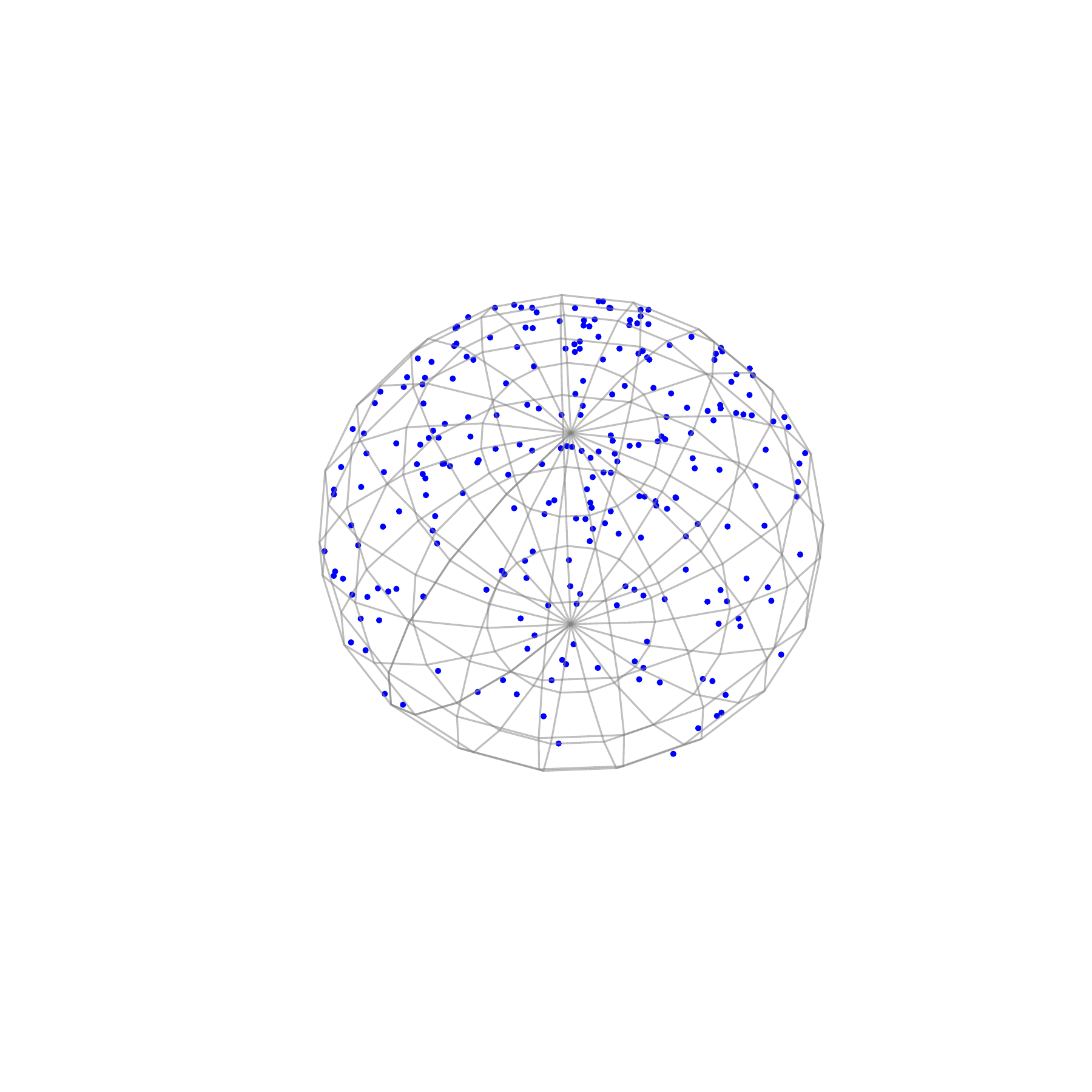}
      }
    \hfill
    \subfloat{
    \includegraphics[width=.48\linewidth,clip=true,trim=100 150 100 200]{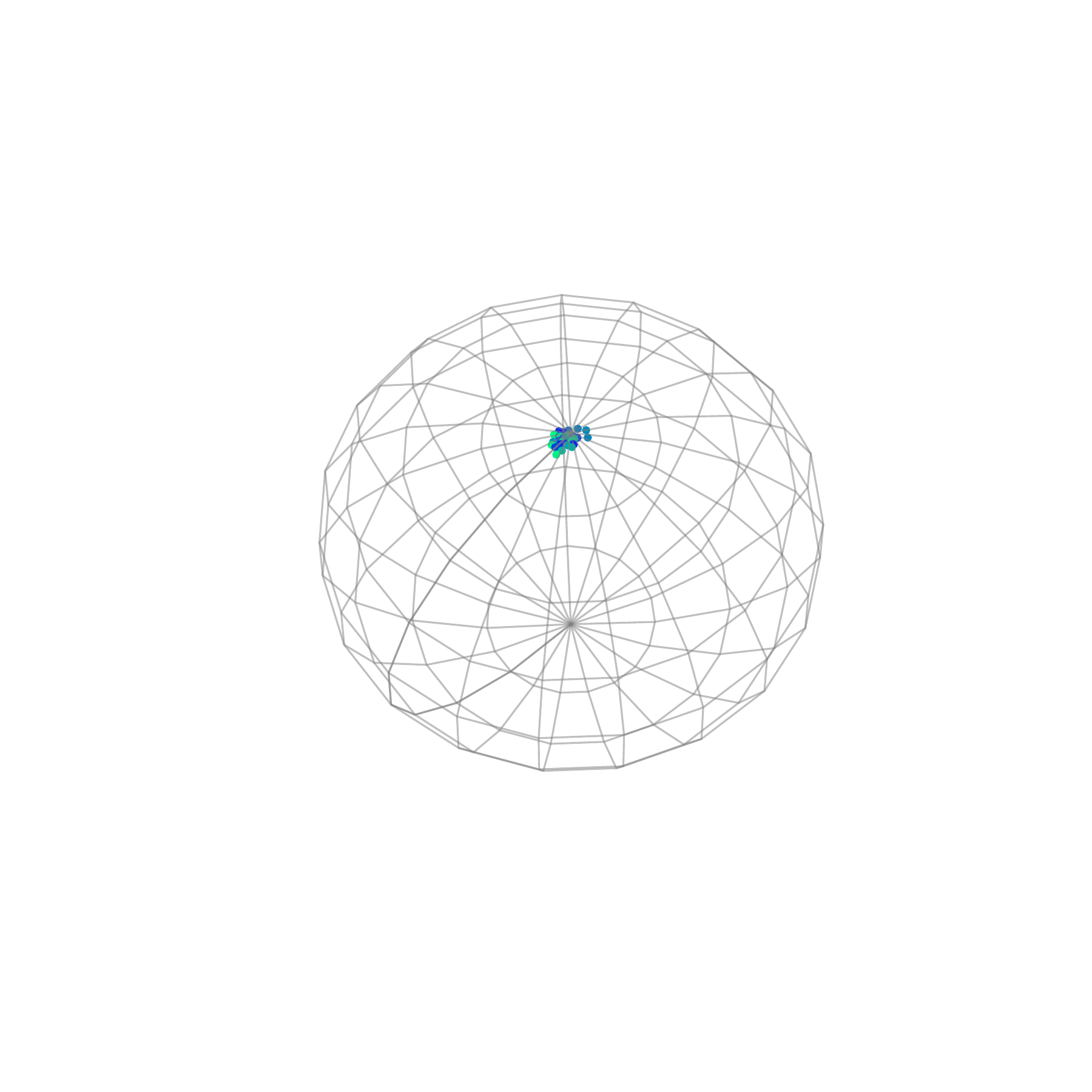}
  }
    \caption{(left) 256 sampled data points on $\mathbb S^2$ (north pole being population mean). (right) 32 samples of the diffusion mean conditioned on the diagonal of $(\mathbb S^2)^n$, $n=256$, $T=.2$. As can be seen, the variation in the mean samples is limited.}\label{fig:s2-2}
\end{figure}

The code used for the experiments is available in the software package
Jax Geometry\footnote{\url{http://bitbucket.org/stefansommer/jaxgeometry}}.
The implementation uses automatic differentiation libraries extensively for the geometry computations as is further described \cite{kuhnel_differential_2019}.

\subsection{Mean estimation on $\mathbb{S}^2$}
To illustrate the diagonal sampling scheme, Figure~\ref{fig:s2-1} displays a sample from a diagonally conditioned Brownian motion on $(\mathbb S^2)^n$, $n=3$. The figure shows both the diagonal sample (red point) and the product process starting at the three data points and ending at the diagonal. In Figure~\ref{fig:s2-2}, we increase the number of samples to $n=256$ and sample 32 mean samples $(T=.2)$. The population mean is the north pole, and the samples can be seen to cluster closely around the population mean with little variation in the mean samples.
\begin{figure}[t]
     \centering
     \subfloat{
        \includegraphics[width=.48\linewidth,clip=true,trim=0 50 0 50]{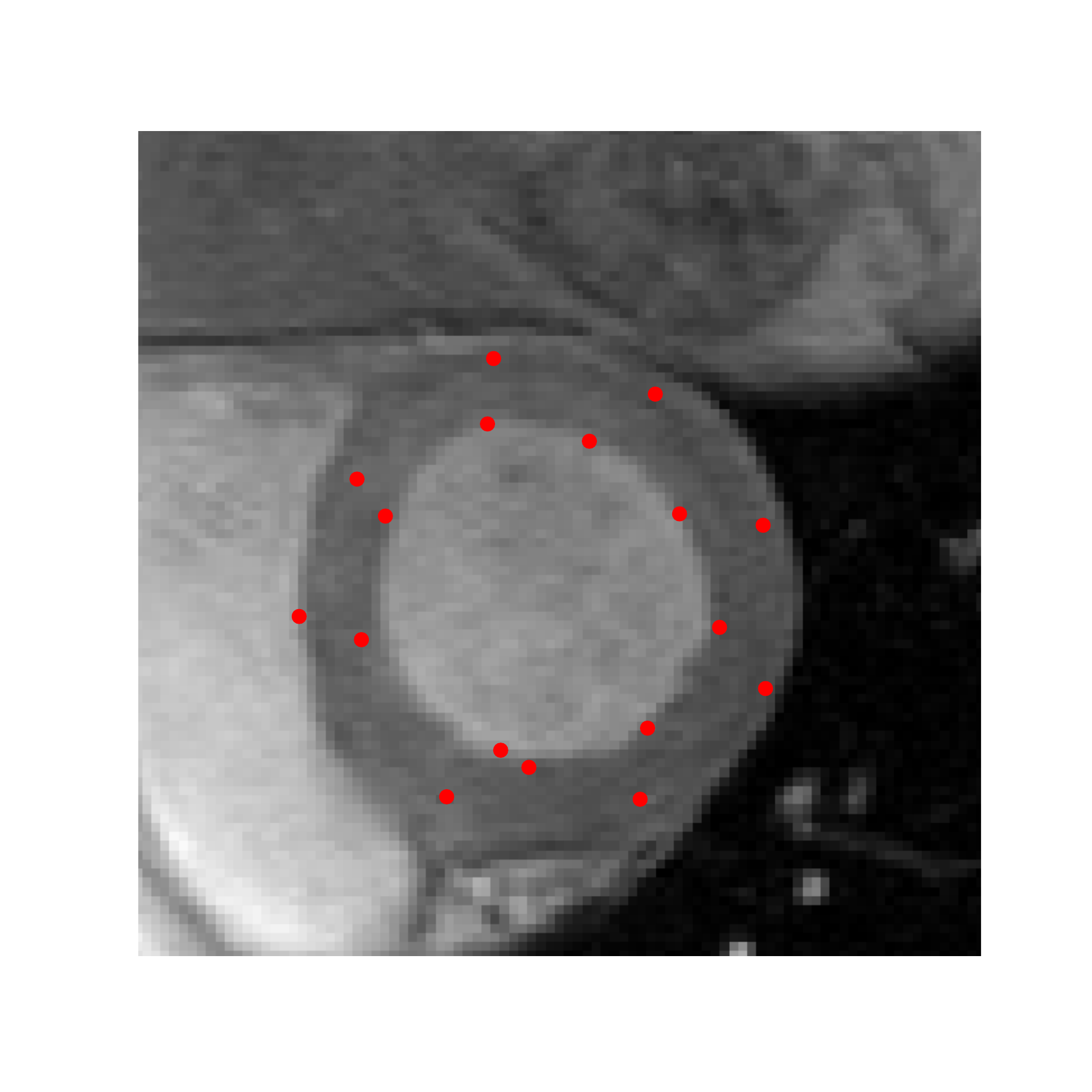}
      }
    \hfill
    \subfloat{
    \includegraphics[width=.48\linewidth,clip=true,trim=0 50 0 50]{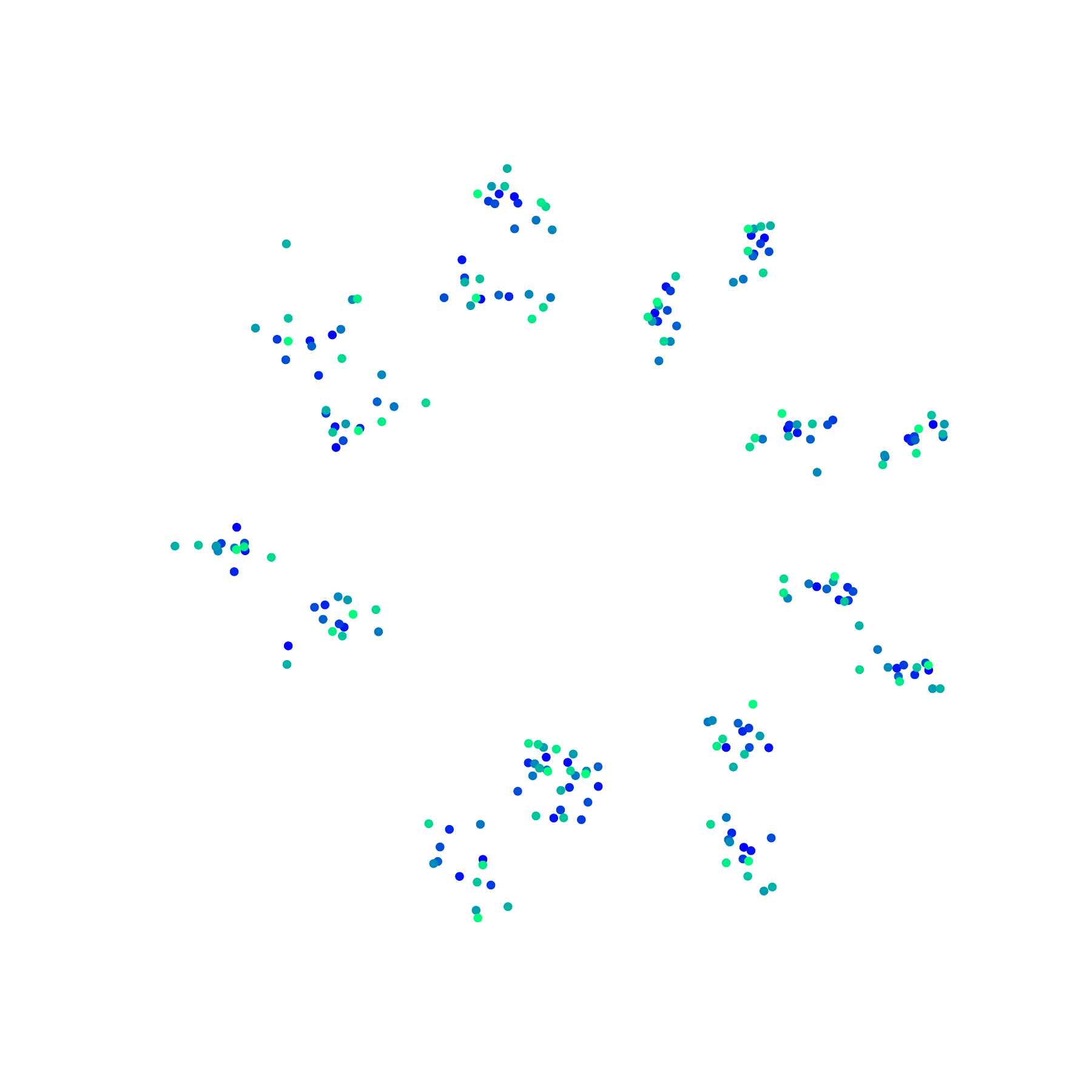}
  }
    \caption{(left) One configuration of 17 landmarks overlayed the MR image from which the configuration was annotated. (right) All 14 landmark configurations plotted together (one color for each configuration of 17 landmarks)}\label{fig:landmarks-1}
\end{figure}

\begin{figure}[t]
     \centering
    \subfloat{
        \includegraphics[width=.48\linewidth,clip=true,trim=0 50 0 50]{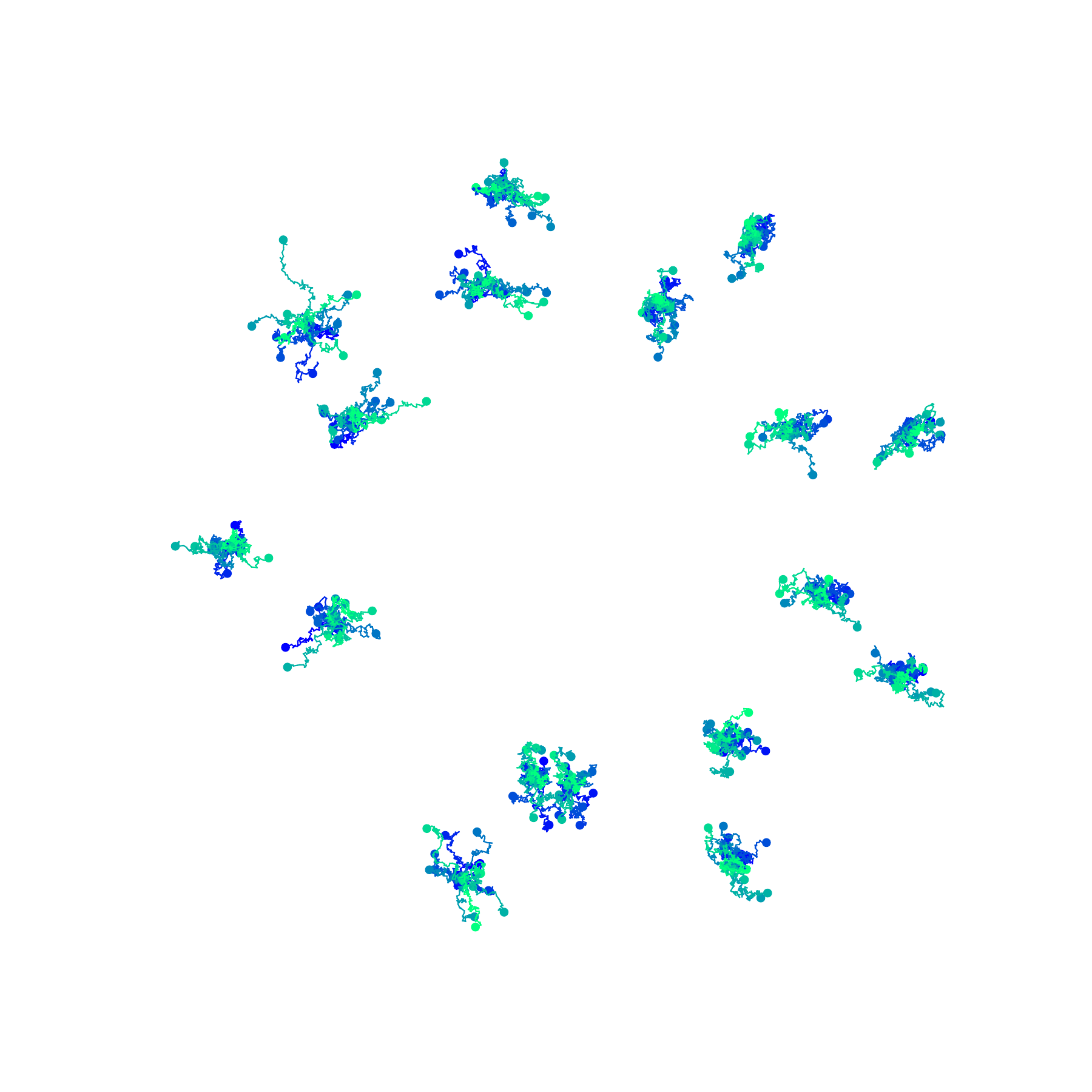}
      }
    \hfill
    \subfloat{
    \includegraphics[width=.48\linewidth,clip=true,trim=0 50 0 50]{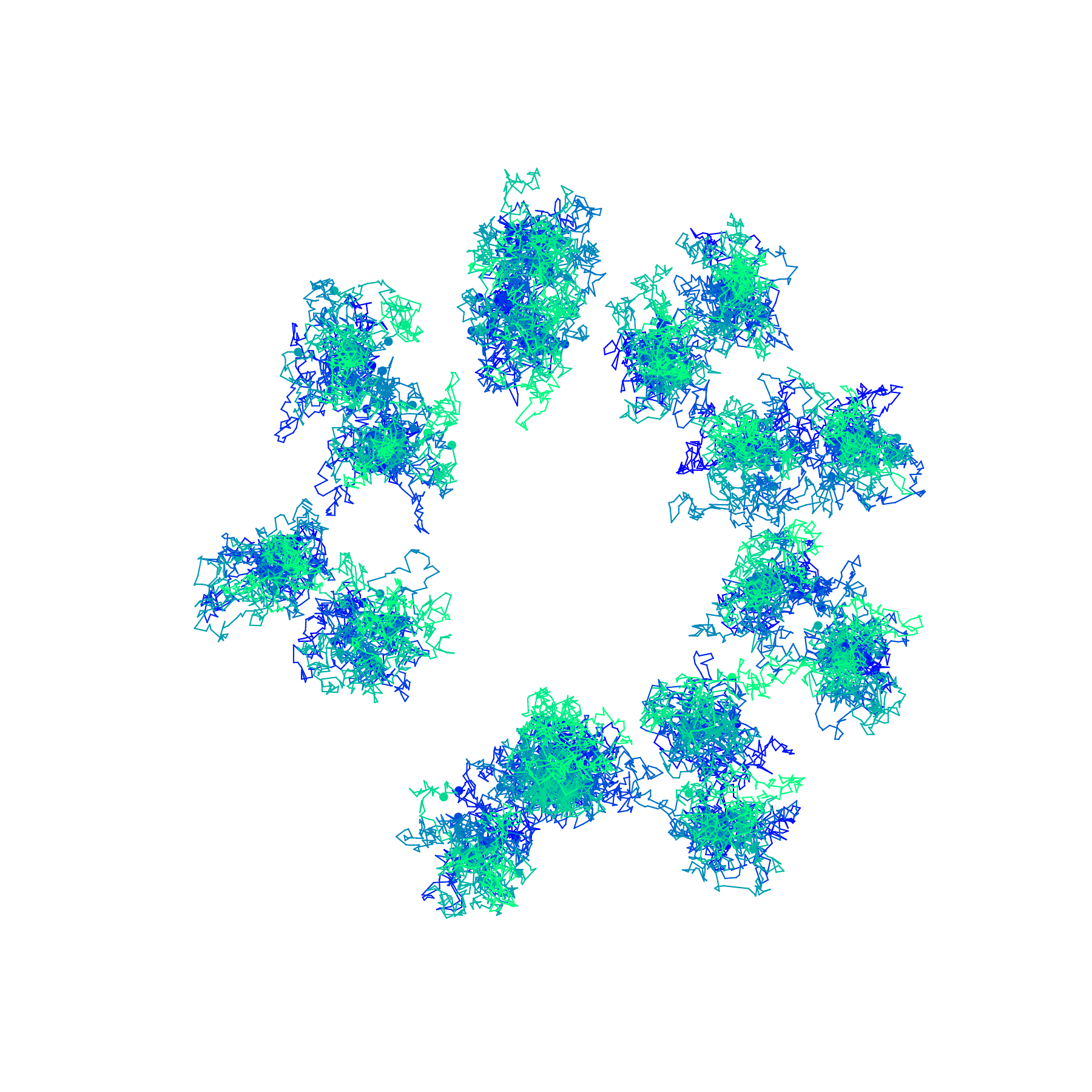}
  }
    \caption{Samples from the diagonal process with $T=.2$ (left) and $T=1$ (right). The effect of varying the Brownian motion end time $T$ is clearly visible.}\label{fig:landmarks-2}
\end{figure}
\begin{figure}[t]
     \centering
    \centering
    \includegraphics[width=.70\linewidth,clip=true,trim=0 50 0 50]{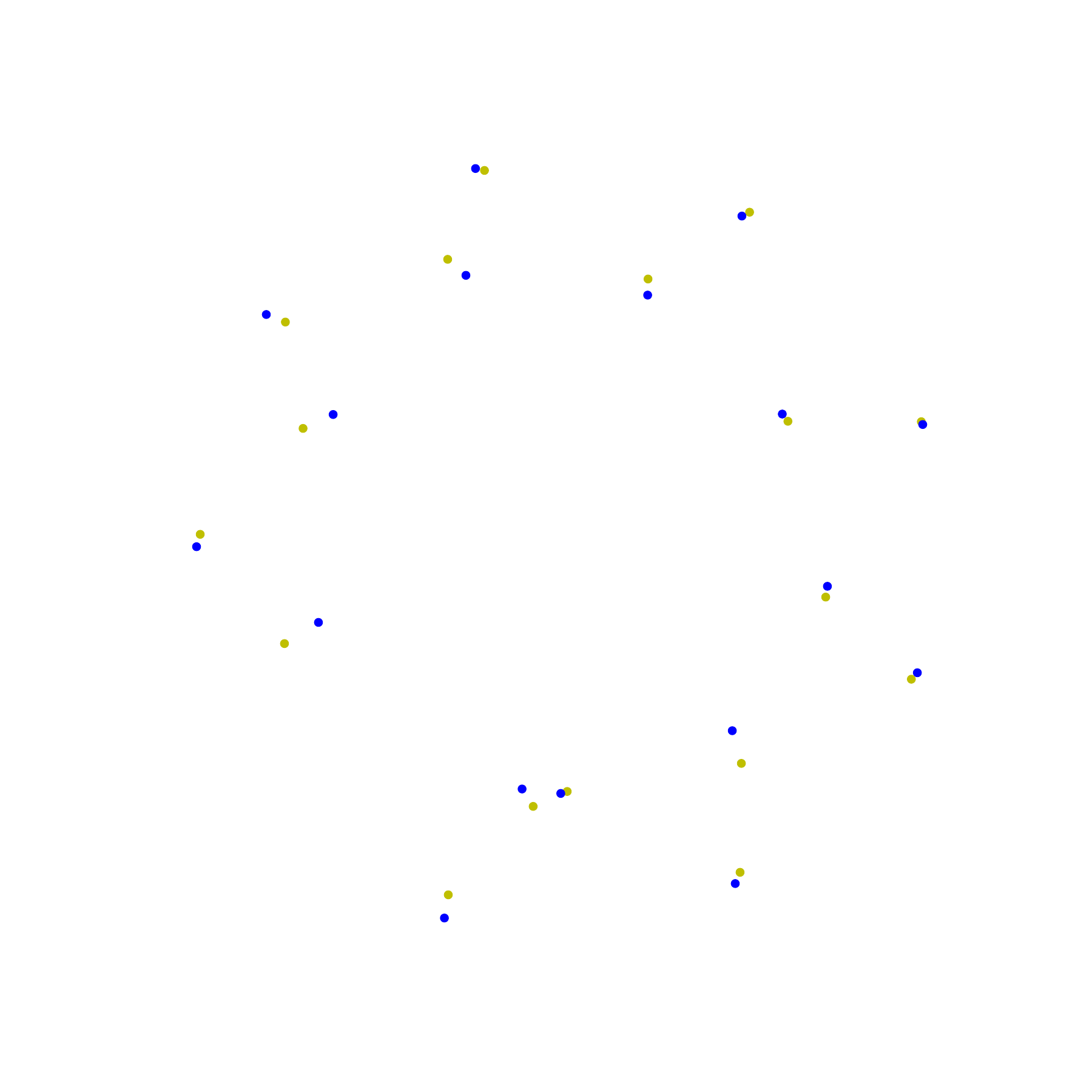}
    \caption{One sampled diffusion mean with the sampling scheme (blue configuration) together with estimated Fr\'echet mean (green configuration). The forward sampling scheme is significantly faster than the iterative optimization needed for the Fr\'echet mean on the landmark manifold where closed form solution of the geodesic equations are not available.}\label{fig:landmarks-3}
\end{figure}

\subsection{LDDMM landmarks}
We here use the same setup as in \cite{sommer_bridge_2017}, where the diffusion mean is estimated by iterative optimization, to exemplify the mean estimation on a high dimensional manifold. The data consists of annotations of left ventricles cardiac MR images \cite{stegmann_extending_2001} with 17 landmarks selected from the annotation set from a total of 14 images. Each configuration of 17 landmarks in $\mathbb R^2$ gives a point in a 34 dimensional shape manifold. We equip this manifold with the LDDMM Riemannian metric \cite{joshi_landmark_2000,younes_shapes_2010}. Note that the configurations can be represented as points in $\mathbb R^{34}$, and the entire shape manifold is the subset of $\mathbb R^{34}$ where no two landmarks coincide. This provides a convenient Euclidean representation of the landmarks. The cometric tensor is not bounded in this representation, and we therefore cannot directly apply the results of the previous sections. We can nevertheless explore the mean simulation scheme experimentally.

Figure~\ref{fig:landmarks-1} shows one landmark configuration overlayed the MR image from which the configuration was annotated, and all 14 landmark configurations plotted together.
Figure~\ref{fig:landmarks-2} displays samples from the diagonal process for two values of the Brownian motion end time $T$. Note that each landmark configuration is one point on the 34 dimensional shape manifold, and each of the paths displayed is therefore a visualization of a Brownian path on this manifold. This figure and Figure~\ref{fig:s2-1} both show diagonal processes, but on two different manifolds.

In Figure~\ref{fig:landmarks-3}, an estimated diffusion mean and Fr\'echet mean for the landmark configurations are plotted together. On a standard laptop, generation of one sample diffusion mean takes approximately 1 second. For comparison, estimation of the Fr\'echet mean with the standard nested optimization approach using the Riemannian logarithm map as implemented in Jax Geometry takes approximately 4 minutes. The diffusion mean estimation performed in \cite{sommer_bridge_2017} using direct optimization of the likelihood approximation with bridge sampling from the mean candidate to each data point is comparable in complexity to the Fr\'echet mean computation.

\section{Conclusion}
In \cite{sommer2020horizontal}, the idea of sampling means by conditioning on the diagonal of product manifolds was first described and the bridge sampling construction sketched. In the present paper, we have provided a comprehensive account of the background for the idea, including the relation between the (weighted) Fr\'echet and diffusion means, and the foundations in both geometry and stochastic analysis. We have constructed two simulation schemes and demonstrated the method on both low and a high-dimensional manifolds, the sphere $\mathbb{S}^2$ and the LDDMM landmark manifold, respectively. The experiments show the feasibility of the method and indicate the potential high reduction in computation time compared to computing means with iterative optimization.

\section*{Acknowledgement} 
The work presented is supported by the CSGB Centre for Stochastic Geometry and Advanced Bioimaging funded by a grant from the Villum foundation, the Villum Foundation grants 22924 and 40582, and the Novo Nordisk Foundation grant NNF18OC0052000.

\bibliography{bibfile,library}

\end{document}